\documentclass[final,3p,twocolumn]{elsarticle}
\biboptions{sort&compress}
\usepackage{amsmath}
\usepackage{amsfonts}
\usepackage{amssymb}

\begin{document}
\title{Consistent description of UCN transport properties}
\author[frmii]{S. Wlokka\corref{cor1}}
\ead{stephan.wlokka@tum.de}
\cortext[cor1]{Corresponding author; Tel.: +49 89 289 13739}
\author[e18]{P. Fierlinger}
\author[frmii]{A. Frei}
\author[ill]{P. Geltenbort}
\author[e18]{S. Paul}
\author[e18]{T. P\"oschl}
\author[e18]{F. Schmid\fnref{fn1}}
\author[e18]{W. Schreyer}
\author[e18]{D. Steffen}
\address[frmii]{Heinz Maier-Leibnitz Zentrum, Technical University of Munich, Lichtenbergstr. 1, D-85748 Garching}
\address[e18]{Physics Department E18 and Universe-Cluster, Technical University of Munich, James-Franck-Str. 1, D-85748 Garching}
\address[ill]{Institut Laue-Langevin, 71 Avenue des Martyrs, 38000 Grenoble, France}
\fntext[fn1]{Present address: Max-Planck-Institut f\"ur Quantenoptik, Hans-Kopfermann-Str. 1, D-85748 Garching}

\begin{abstract}
We have investigated the diffuse reflection probabilities of Replica guides for ultra-cold neutrons (UCN) using the so-called helium method. For the first time we could establish a consistent description of the diffuse reflection mechanism for different lengths of the guide system. The transmission of the guides is measured depending on the helium pressure inside of the guides. A series of simulations was done to reproduce the experimental data. These simulations showed that a diffuse reflection probability of $d = \left( 3.0 \pm 0.5 \right) \cdot 10^{-2}$ sufficiently describes the experimental data. 
\end{abstract}
\maketitle
\section{Introduction}
Ultra-cold neutrons (UCN) \cite{golub1975, ignatovich1990} are neutrons, which are reflected from a material surface under all angles of incident due to the Fermi potential \cite{fermi1936} of the material. Typically they have energies lower than $300 \, \mathrm{neV}$, which corresponds to velocities smaller than $8 \, \mathrm{m/s}$. UCN can be stored in material or magnetic bottles for long times (several hundreds of seconds). Thus one can determine fundamental properties of the neutron itself such as its lifetime \cite{picker2005} or its electric dipole moment (EDM) \cite{altarev2010}.

Currently new sources for ultra-cold neutrons are being built. \cite{korobkina2007, lauss2011, martin2008,frei2007,frei2014}. Efficient transport of ultra-cold neutrons requires guides with a high fermi-potential, low absorption and a low diffuse reflection probability. The latter describes the probability for UCN to be reflected non-specularly. This essentially allows backscattering of the UCN and strongly increases the loss probability. In order to ensure the delivery of a high amount of UCN to the experiments at the Forschungs-Neutronenquelle Heinz Maier-Leibnitz (FRM II) we developed UCN guides with relative transmissions of $\left( 0.990 \pm 0.006 \right) \, \mathrm{m^{-1}}$ \cite{huber2011}. These guides are made of a Ni-V alloy and are produced by the so-called ``Replika''-technique \cite{steyerl1986,plonka2007}. In the course of the described experiment we further investigated the transmission properties focussing on the diffuse reflectivity applying the helium method.

\section{The Helium Method}
Beginning with the stationary diffusion equation
\begin{equation}
\frac{\mathrm{d}^2 n}{\mathrm{d} z^2} = \frac{n}{L_\mathrm{D}^2},
\label{eq_diffusion_equation}
\end{equation}
$n$ being the neutron density along the length of the guide axis $z$, one can deduct a solution for the neutron flux $J$ at a given position $L$ within the guide \cite{groshev1971, ignatovich1996}.
\begin{equation}
J \! \left( L \right) = \frac{J \! \left( 0 \right)}{\left( \mathrm{cosh} \! \left( \frac{L}{L_\mathrm{D}} \right) + \gamma \, \mathrm{sinh} \! \left( \frac{L}{L_\mathrm{D}} \right) \right)}.
\label{eq_neutron_flux}
\end{equation}

\begin{figure*}[ht]
\centering
\includegraphics[width=0.8\textwidth]{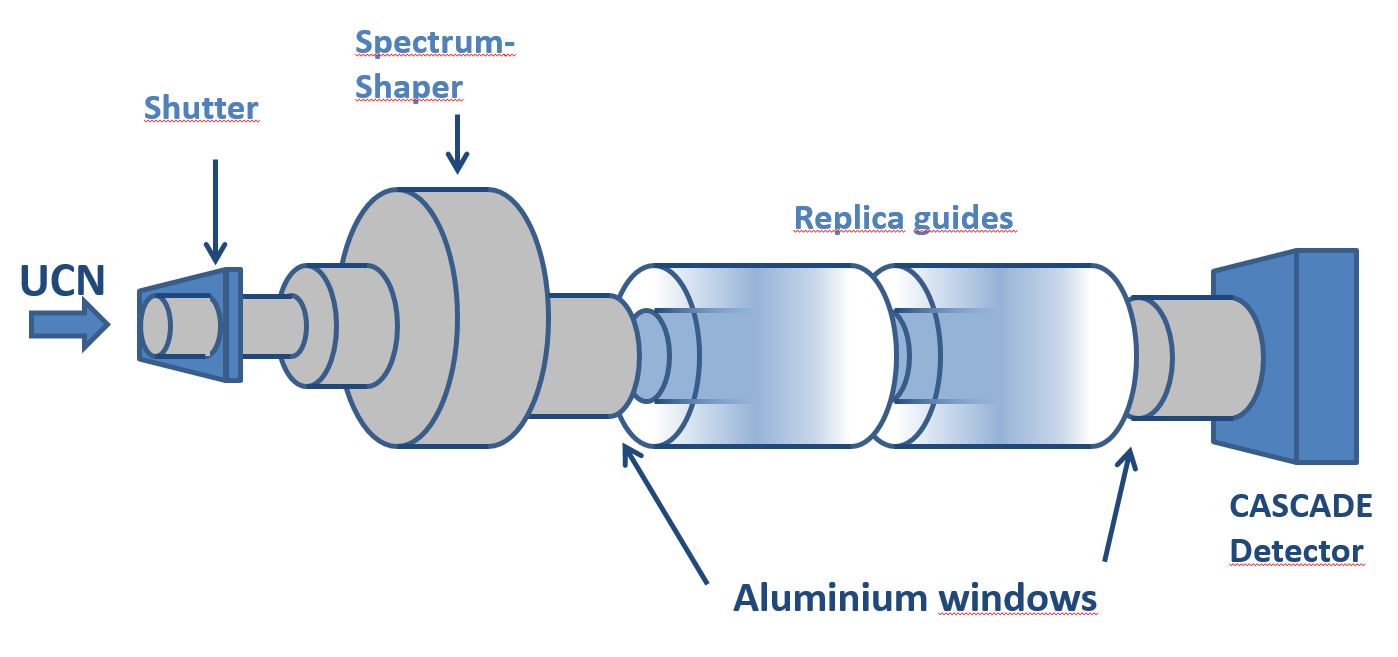}
\caption{(Color online) Experimental setup used for the helium method. UCN enter from the left. A shutter is used to open and close the Replika volume for UCN. After passing through a spectrum shaper \cite{daum2012} the neutrons enter the Replika volume, which is enclosed by Aluminium foils on both ends. A Cascade Detector \cite{klein2014} is used for the UCN detection.}
\label{figure_exp_setup}
\end{figure*}

Here, $L_\mathrm{D} = \sqrt{D \tau}$ denotes the diffusion length with the diffusion constant $D$ and the effective lifetime of the neutron in the guide system $\tau$ and $\gamma = \left( L_\mathrm{D} \bar{v} \right) / \left( 4 D \right)$ with the mean neutron velocity $\bar{v}$.

Diffuse reflections in the guide system decrease the diffusion constant $D$ and thus the diffusion length $L_\mathrm{D}$ effectively reducing the transmission of the guides. There is still some debate about the physics behind the diffuse reflection mechanisms for UCN \cite{steyerl2010, serebrov2010}. In our analysis we will employ the Lambert model where the neutrons have a cosine distribution about the normal vector of the scattering surface.

Measurements of the diffuse reflectivity usually employ a slit and collimator system, which reduces the angles accepted by the guide system to a narrow band of a few degrees. \cite{atchison2010} In our case adding helium to the system decreases the lifetime of the neutrons in the guide. Since the helium atoms are much faster ($v_\mathrm{RMS} \sim 1260 \, \mathrm{m/s}$) than the UCN, any interaction of the neutrons with the gas atoms lead to an upscattering of the UCN and thus to the loss of it. As a higher probability for diffuse reflection increases the path length a neutron has to take through the guide system, the loss probability in the helium gas is also increased.

Generally one could follow the suggestion of \cite{ignatovich1996} and measure the helium pressure needed to decrease the UCN count rate to $1/e$ with respect to vacuum. However, in our setup (cf. Section \ref{sec_setup}) the UCN spectrum differs severely from the spectrum assumed in \cite{ignatovich1996} both in the energy aswell as in the momentum distribution. Therefore we do not use Eq. \ref{eq_neutron_flux} to extract the diffuse reflectivity but we rely on simulations, which will be described in Section \ref{sec_simulation}.

\section{Experimental Setup}
\label{sec_setup}
Figure \ref{figure_exp_setup} shows the experimental setup of the experiment. UCN from the PF2/TES beamline \cite{steyerl1986} at the Institute Laue-Langevin enter the experiment from the left. A shutter at entrance of the beam tube controls the flux into the experiment. UCN then pass a stainless steel spectrum shaper, which cuts off most neutrons above $\sim 190 \, \mathrm{neV}$ \cite{daum2012}. Downstream the neutrons enter the Replika guide volume, which is enclosed on both sides by aluminum (AlMg3) windows with a thickness of $100 \, \mathrm{\mu m}$. After the UCN passed the second window they are guided to a standard UCN detector of the CASCADE type \cite{klein2014}. The detector signals were amplified with a CAEN 968 Amplifier and analysed in the PC by a MCA 3A.

The volume between the two aluminum windows could be filled with helium with a purity of $99.996 \, \mathrm{\%}$. Before the helium was put into our system it was flushed through a liquid nitrogen bath to eliminate possible contaminants (such as water).

The length of Replika section was varied between $1-3 \, \mathrm{m}$. Every experimental cycle began with a transmission measurement under vacuum conditions. Afterwards, helium was filled into the Replika volume up to pressures ranging between $10 - 500 \, \mathrm{mbar}$. The count rate at the detector for different pressures values were normalized to the count rate at zero pressure after correcting the data for the time-varying reactor power.

\begin{figure}[!b]
\centering
\includegraphics[width=0.45\textwidth]{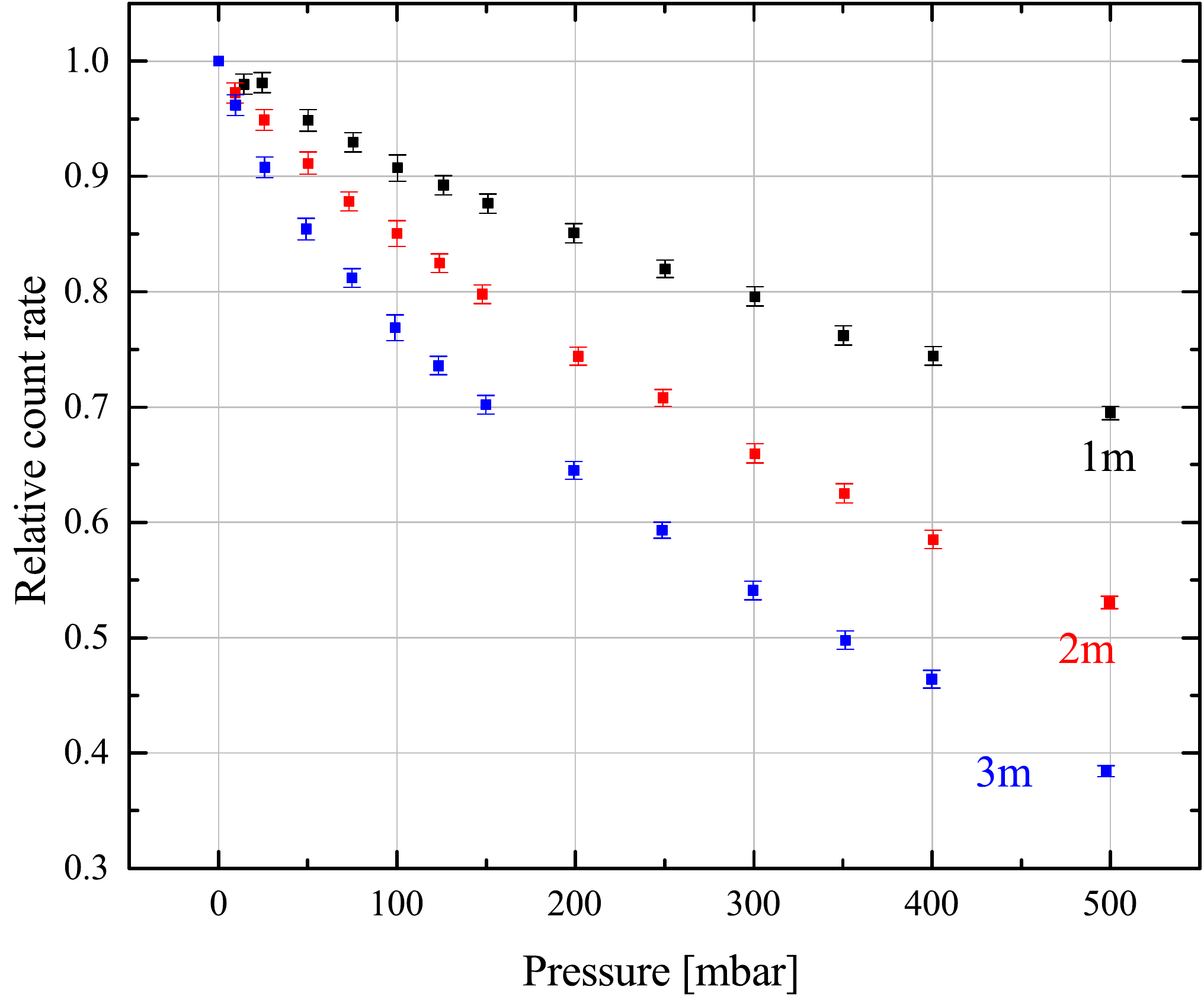}
\caption{(Color online) Normalized count rates for different pressures using Replika guide lengths between $1-3 \, \mathrm{m}$. The error bars are statistical uncertainties plus uncertainties from a reactor power correction. All count rates are shown relative to the vacuum count rate.}
\label{figure_helium_curves}
\end{figure}

\section{Results}
\label{sec_results}
Figure \ref{figure_helium_curves} shows the results of our experiments for three different lengths of Replika guides. All values are normalized to the respective measurements at vacuum conditions. With increasing pressure the transmission decreases owing to an increased interaction probability of UCN and helium. As expected, the transmission decreases for the longer guides.

\begin{table*}[t]
\caption{Calculated values for the mean path lengths of UCN.}
\begin{center}
\begin{tabular}{r || c | c | c}
\centering
Guide length	&	$L_1 [\mathrm{m/m}]$	&	$L_2 [\mathrm{m/m}]$	&	$\bar{L} [\mathrm{m/m}]$ \\
\hline
\hline
$1 \, \mathrm{m}$	&	$1.47 \pm 0.04$	&	$35.60 \pm 13.49$	&	$2.66 \pm 0.53$ \\
\hline
$2 \, \mathrm{m}$	&	$1.22 \pm 0.03$	&	$11.46 \pm 1.29$	&	$2.33 \pm 0.17$ \\
\hline
$3 \, \mathrm{m}$	&	$1.28 \pm 0.01$	&	$23.43 \pm 2.65$	&	$3.30 \pm 0.26$ 
\end{tabular}
\end{center}
\label{table_length_values}
\end{table*}

Considering the case of no diffuse reflectivity and a gaussian UCN energy distribution, one would expect the data to follow an exponential distribution

\begin{equation}
\frac{I}{I_0} = e^{-n_\mathrm{He}\left( \sigma_\mathrm{sc} + \sigma_\mathrm{abs} \right) \bar{L}}
\label{eq_exponential_transmission}
\end{equation}

with the helium atom density $n_\mathrm{He}$, the total scattering cross-section $\sigma_\mathrm{sc}$, the absorption cross-section $\sigma_\mathrm{abs}$ and the mean path length $\bar{L}$. The mean UCN energy can be extracted from our simulation to be $\bar{E}_\mathrm{UCN}=185\,\mathrm{neV}$ ($\bar{v} = 5.95 \, \mathrm{m/s}$).

\begin{figure}[!b]
\centering
\includegraphics[width=0.45\textwidth]{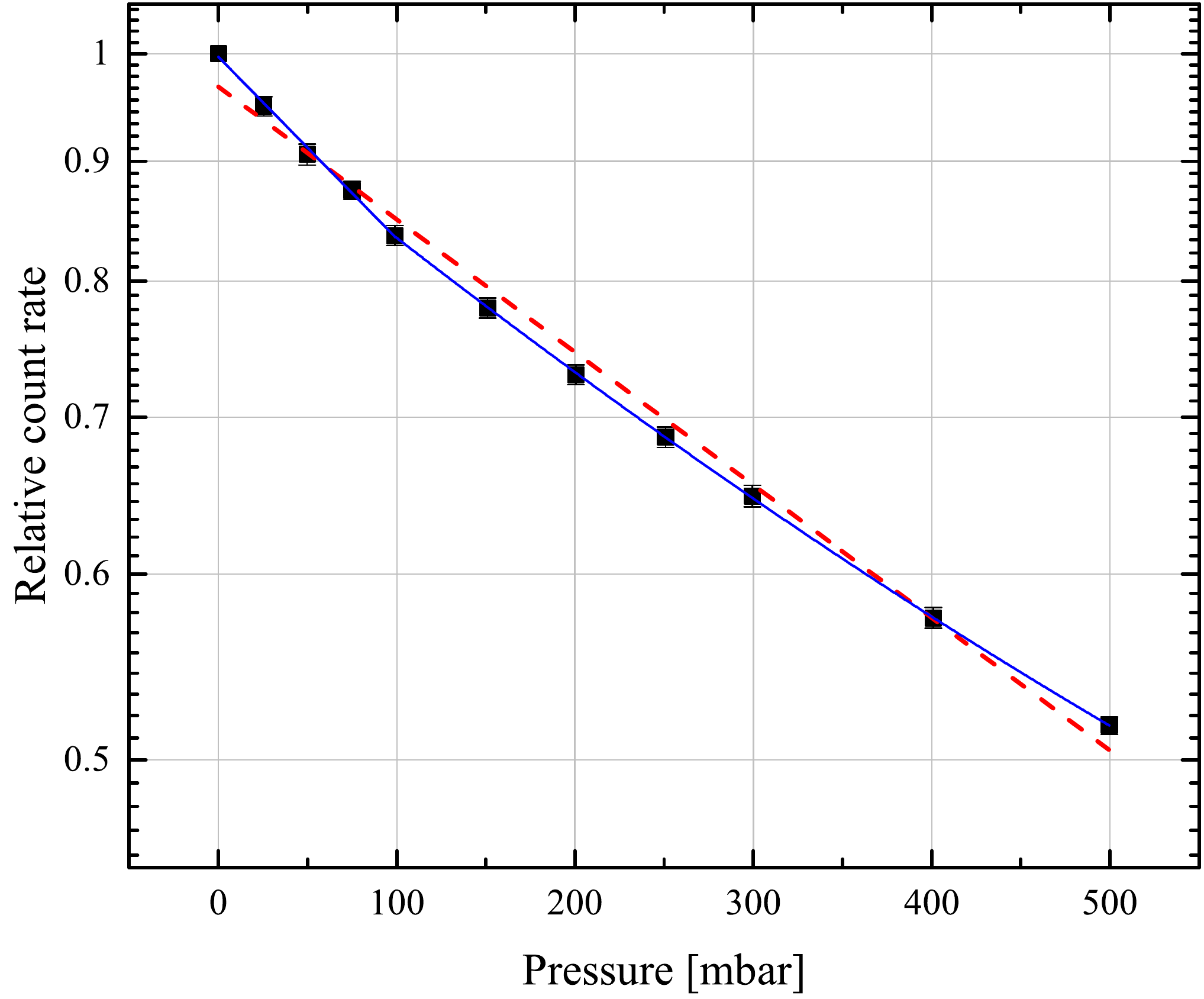}
\caption{(Color online) Normalized UCN count rate for $L_\mathrm{Replika} = 2 \, \mathrm{m}$ as function of the He pressure in logarithmic scale. The solid line represents a fit using two exponentials. For comparison we also show the best single exponential fit in a dashed line.}
\label{figure_2m_exponentials}
\end{figure}

From fig. \ref{figure_2m_exponentials} one can see that the helium curves in fact do not have a true exponential form indicated by the red line. The blue line shows two separate exponentials. For low diffusely reflecting guides one would expect the UCN to undergo very few diffuse scattering events. Considering a binomial distribution of the diffuse scattering events one can estimate that there are less than five diffuse scattering events through the tested guides given a diffuse reflection probability of $2 \, \mathrm{\%}$.

Our measurements show that the UCN sample detected at the exit consists of two different populations. One population has undergone a large number of surface reflections, which leads to an increased path length and thus high loss rates in He already at low pressures. The other population is characterized by a direct path through the guide system. This behaviour is visible for all three lengths between $100-150 \, \mathrm{mbar}$, however, it is more prominent for the longest guide length since the probability for zero diffuse scattering events is reduced.

Using the mean energy from our simulations, one can calculate the mean path lenghts from the double exponential fit. The results are summarized in table \ref{table_length_values}. The short characteristic path lengths $L_1$ agree with the mean path length obtained from our simulations assuming no diffuse reflectivity ($1.27 \, \mathrm{m/m}$). The longer path length $L_1$ for the short Replika piece is due to a change in the momentum spectrum shortly after leaving the spectrum shaper. The lengths $L_2$, corresponding to UCN scattered away from the forward direction, are an order of magnitude longer than $L_1$. The weighted mean $\bar{L}$ (using the factors from the exponential fit as weights) does correspond well with the mean path length extracted from our simulation with diffuse reflection ($2.22 \, \mathrm{m}$).

\section{Simulation}
\label{sec_simulation}
In order to extract values for the diffuse reflectivity of the Replika guides, a \textsc{GEANT4} \cite{agostinelli2003} simulation was used. This simulation incorporated all parts of the experimental setup. The simulated velocity spectrum had a gaussian shape with a peak at $10 \, \mathrm{m/s}$ and a spread of $5 \, \mathrm{m/s}$ roughly corresponding to the situation at the PF2/TES beam line \cite{zechlau2015}. The velocity vectors were generated following a cosine distribution around the guide axis. The simulation incorporated the TES beam aluminum exit window as well as our shaper.

The following expression was used to describe the interaction cross section between UCN and the helium gas \cite{beckurts1964}:
\begin{equation}
\sigma_\mathrm{up} \! \left( E_\mathrm{UCN} \right) = 2 \sigma_\mathrm{0} \sqrt{\frac{k_\mathrm{B}T}{A \pi E_\mathrm{UCN}}} \times \frac{4 m_\mathrm{n} M}{\left( M + m_\mathrm{n} \right)^2}.
\label{eq_liu_helium_crosssection}
\end{equation}
Here, $\sigma_\mathrm{0}$ is the thermal scattering cross-section ($1.34 \, \mathrm{barn}$ in the case of helium), $A = 4$ is the mass ratio of helium atoms and a neutrons, $m_\mathrm{n}$ and $M$ are the masses of neutrons and helium atoms respectively and $E_\mathrm{UCN}$ is the energy of the UCN. Any interaction of a UCN with a helium atom leads to the UCN being upscattered to thermal energies. Thus, in the simulation, the UCN, which interacted with the helium were removed from the ensemble.

To describe diffuse scattering of the neutrons on the walls, we used a modified Lambertian reflection mechanism, where the neutrons are scattered with a cosine law around the axis of specular reflection instead of an axis normal to the surface.

In the simulations we also assumed the aluminum foils to be covered by a layer of Fombline oil due to a contamination of the facility by previous experiments, which in turn alters the Fermi potential of the foils (i.e. $V_\mathrm{Fom} = 106\,\mathrm{neV}$, $V_\mathrm{Al} = 54\,\mathrm{neV}$). Thus, the energy spectrum of UCN at the entrance of the Replika volume is significantly altered, especially their angular distribution, which is much more forward peaked than for pure aluminum and consequently enhances the relative transmission.

\begin{figure}
\centering
\includegraphics[width=0.45\textwidth]{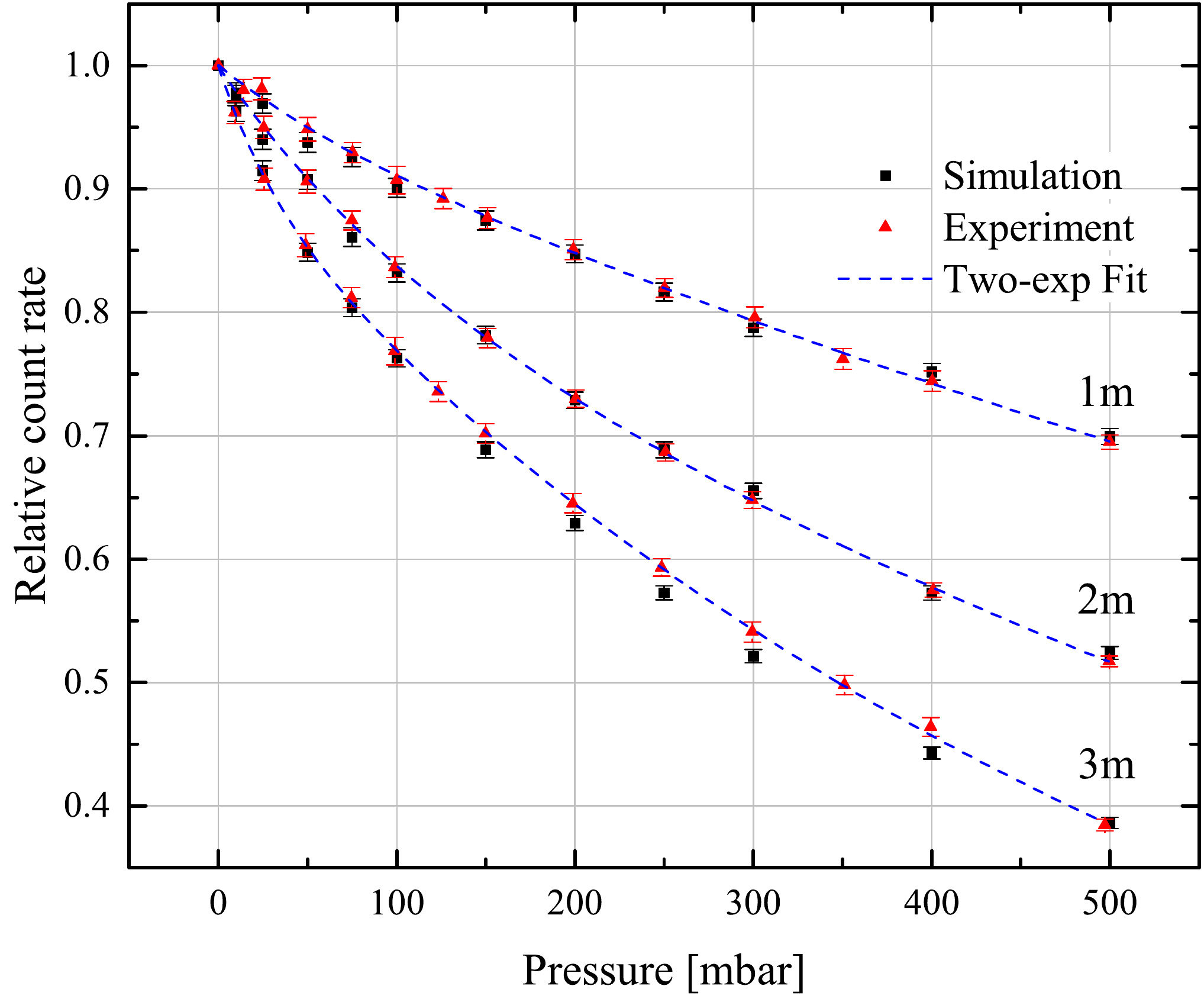}
\caption{(Color online) Simulated (squares) and experimental (triangle) data. The experimental data are shown with the respective double exponential fits (dashed line). Counts are relative to the count rates with no helium in the simulation.}
\label{figure_helium_curves_comparison}
\end{figure}

Fig. \ref{figure_helium_curves_comparison} shows the simulation together with the experimental data. The diffuse reflectivity of the Replika pieces in the simulation was set to $d = 3.0 \cdot 10^{-2}$. Simulated and experimental data match well. We define an error number $\epsilon^2$ number in the following way

\begin{equation}
\epsilon^2 = \frac{\left( y_\mathrm{dat} - y_\mathrm{sim} \right)^2}{\sigma^2_\mathrm{dat} + \sigma^2_\mathrm{sim}}
\label{eq_chi_sq_function}
\end{equation}

with the relative counts for the real data and simulation ($y_\mathrm{dat}$, $y_\mathrm{sim}$) and their respective errors. We receive a $\epsilon^2/\mathrm{DOF}$-values of $0.91$. By varying the diffuse reflectivity in the simulation we determine the uncertainty of the value for the diffuse reflectivity to be $\sigma_d = 0.5 \cdot 10^{-2}$.

The measurements are described consistently and our model could be applied to different lengths of a given guide material with a unique set of parameters in the simulation. We thus conclude that the modified Lambert model is the correct description for diffuse scattering on very smooth surfaces.
\section{Conclusions}
Guides for ultra-cold neutrons made from a NiV alloy have been studied before with a different method \cite{atchison2010}. Atchison et. al obtained values for the diffuse reflection probability of around $0.01$ for their samples. However, this value has been extracted using the original Lambertian scattering mechanism. Under these circumstances our results are comparable.

We have successfully applied the helium method to our Replika guides. We determined a diffuse reflectivity of $d = \left( 3.0 \pm 0.5 \right) \cdot 10^{-2}$ using a modified Lambertian scattering mechanism. It is the first time that a UCN guide system could be characterized by a single value for $d$ for different guide lengths. Combining our previous measurements of the NiV Replika guides and this work, we arrive at a good understanding of their transmission properties.

The helium method itself is rather easy to prepare and conduct, thus it provides a cheap means of testing guides for ultra-cold neutrons on their reflection properties.

\section{Acknowledgement}
We thank T. Brenner and T. Deuschle for their support in the preparation of the experimental setup and during the beamtime. We also thank C. Morkel for the helpful discussions during data analysis.\par
This work was funded by the DFG Excellenz-Cluster EXC153 ''Origin and Structure of the Universe'' and the Maier-Leibnitz-Laboratorium (MLL) of the TU and LMU Munich.

\bibliography{heliummethod}
\end{document}